\documentclass[sigconf, nonacm]{acmart}

\newcommand\vldbdoi{XX.XX/XXX.XX}

\newcommand\vldbavailabilityurl{URL_TO_YOUR_ARTIFACTS}
\newcommand\vldbpagestyle{plain} 

\usepackage{minted}
\usepackage{graphicx}
\usepackage{dirtytalk}
\usepackage{amsmath,empheq}
\usepackage{mathrsfs}
\usepackage{tikz}
\usepackage{neuralnetwork}
\usepackage{subfig}
\usepackage{booktabs}
\usepackage[ruled,vlined,linesnumbered]{algorithm2e}
\usepackage{setspace}
\usepackage{xcolor}
\usepackage{url}
\usepackage{enumerate}
\usepackage{diagbox}
\usepackage{rotating}
\usepackage{float}
\usepackage{hyperref}
\usepackage{comment}
\usepackage{multirow}
\usepackage{nopageno}
\usepackage{enumitem}
\usepackage[normalem]{ulem}
\usepackage{soul}
\usepackage{balance}
\usepackage{pifont}
\usepackage{ifthen}
\usepackage{balance}
\usepackage{pdfpages}

\usepackage{algpseudocode}
\usepackage{threeparttable} 
\usepackage{makecell}

\usepackage{listings}
\usepackage{xcolor}
\usepackage{courier} %

\newtheorem{definition}{Definition}

\lstdefinestyle{mystyle}{
  basicstyle=\ttfamily\small,
  keywordstyle=\color{blue},
  commentstyle=\color{gray},
  stringstyle=\color{orange},
  numbers=left,
  numberstyle=\tiny\color{gray},
  stepnumber=1,
  breaklines=true,
  frame=single,
  captionpos=b,
  language=Python
}
\newcommand{\fix}[1]{{\color{black}#1}}

\newcommand\ourtool{DriftBench}

\begin{document}
\title{DriftBench: Defining and Generating Data and Query Workload Drift for Benchmarking}


\author{Guanli Liu}
\affiliation{%
  \institution{The University of Melbourne}
}
\email{guanli.liu1@unimelb.edu.au}

\author{Renata Borovica-Gajic}
\affiliation{%
  \institution{The University of Melbourne}
}
\email{renata.borovica@unimelb.edu.au}

\renewcommand{\shortauthors}{\emph{et al.}}

\begin{abstract}
Data and workload drift are key to evaluating database components such as caching, cardinality estimation, indexing, and query optimization. Yet, existing benchmarks are static, offering little to no support for modeling drift. This limitation stems from the lack of clear definitions and tools for generating data and workload drift.

Motivated by this gap, we propose a unified taxonomy for data and workload drift, grounded in observations from both academia and industry. Building on this foundation, we introduce DriftBench, a lightweight and extensible framework for generating data and workload drift in benchmark inputs.
Together, the taxonomy and DriftBench provide a standardized vocabulary and mechanism for modeling and generating drift in benchmarking. We demonstrate their effectiveness through case studies involving data drift, workload drift, and drift-aware cardinality estimation.
\end{abstract}

\maketitle

\pagestyle{\vldbpagestyle}



\section{Introduction}\label{sec:introduction}
Have you ever read or reviewed a database paper and wondered: \textit{does this system adapt to data or workload\footnote{For brevity, we use \textit{workload} to refer to \textit{query workload} throughout the paper.} drift\footnote{``Drift'' and ``shift'' are often used interchangeably in data and workload characteristics. 
In this paper, we use \textit{drift}, following the terminology in NeurDB~\cite{NeurDB}.
}?}
Or perhaps you have been on the receiving end of the same question, one that is easy to ask but surprisingly difficult to answer rigorously.

We raise this question because both data and workload drift have become increasingly prevalent in real-world systems~\cite{NeurDB,modeling_shift,workload_shift_detection}. 
As a result, they have attracted growing attention in recent research such as query optimization, cardinality estimation, and learned indices~\cite{Bao, Neo, Balsa,LBMC,DBA_bandit, noDBA, hmab, selep, aulid, ULI_Evaluation, RSMI, ELSI, Poisoning, LMSFC, MSCN, LW-NN}.

However, existing research works lack a unified definition of data and workload drift, as well as principled methods for generating them. 
Instead, they typically approximate drift by switching between multiple static datasets~\cite{sosd,ULI_Evaluation,learnedbench,learned_spatial_indexes,learnedmultiindexNeurIPS,are_learned_ready}, or by manually varying query parameters (e.g., predicate ranges) and query types (e.g., $k$NN and range queries in spatial workloads).
While these heuristics can simulate drift, they 
remain ad hoc and may fall short in capturing the dynamic nature of real-world drift.

This lack of standard definitions and tools is also reflected in existing benchmarks.
For example, TPC-H~\cite{TPCH} and TPC-DS~\cite{TPCDS}  are widely used benchmarks that assume static data and fixed workloads, supporting data scaling via a scale factor but lacking mechanisms to model drift.
The Join Order Benchmark (JOB)~\cite{job} focuses on join queries
but does not support dynamic workloads.
RedBench~\cite{redbench} improves real-world representativeness by sampling production workloads from Amazon Redshift~\cite{TPC_not_enough}, but its fixed samples cannot support controlled drift injection over arbitrary data schemas.
Thus, drift evaluations remain ad hoc, with limited reproducibility and comparability across systems.

In particular, for data drift, no widely accepted definitions capture common forms of change such as distributional drifts and cardinality variations, or even schema modifications~\cite{NeurBench,NeurDB,Warper}. 
The notion of workload drift is even more ambiguous~\cite{modeling_shift,workload_shift_detection}. For example, whether changes in query selectivity, predicate distributions, or query templates constitute workload drift.
These distinctions are rarely formalized, leaving practitioners without clear guidance.

Clarifying the definition of drift is only part of the solution. The community also needs practical tools for constructing drift scenarios across diverse datasets and schemas. Ideally, users should be able to create data drift by modifying distributions and inserting/deleting data records.
Similarly, workload drift should be expressible through controllable changes including predicate distributions, predicate ranges, and query structures.


To fill this gap, we define a taxonomy of data and workload drift.
We then propose \textbf{\ourtool}\footnote{Artifacts available at: \url{https://github.com/Liuguanli/DriftBench}}, a lightweight and extensible framework for tabular datasets that supports three key core capabilities for drift-aware benchmarking.
First, it enables \textbf{data drift} through transformations of real datasets, including varying cardinality, generating insert/delete operations, shifting column distributions, and injecting outliers.
Second, it supports \textbf{workload drift} 
through modifying predicate distributions and ranges in templates, and mutating predicates, join conditions, and payloads.
Third, it generates timestamp sequences exhibiting \textbf{temporal drift patterns}~\cite{Sibyl}, including trending, periodic cycles, and long-tail evolution.


This paper makes the following contributions:
\begin{itemize}[leftmargin=1em, topsep=0pt, itemsep=0pt, parsep=0pt]

    \item We formalize data and workload drift into four representative operations each, along with four temporal drift patterns, providing a clear taxonomy for drift-aware benchmarking.

    \item We present \ourtool, a lightweight and extensible framework that generates drifted data and workloads from tabular sources, e.g., CSV files and PostgreSQL.

    \item We demonstrate how \ourtool\ generates data and workload drift with temporal patterns using a real-world dataset, and showcase its practical applicability through a drift-aware cardinality estimation case study.

\end{itemize}

    

\begin{table}[h]
\small
\centering
\renewcommand{\arraystretch}{1.2}
\begin{tabular}{|p{1.2cm}|p{1.6cm}|p{4.4cm}|}
\hline
\textbf{Drift} & \textbf{Subtype} & \textbf{Operation} \\
\hline
\multirow{4}{*}{Data} 
    & \textit{Cardinality}     & \textbf{Varying Cardinality} \\
    \cline{3-3}
    &                & \textbf{Updating Cardinality} \\
\cline{2-3}
\cline{2-3}
    & \textit{Distributional} & \textbf{Shifting Column Distributions}\\
    \cline{3-3}
    &                & \textbf{Injecting Outliers}\\
    
\hline
\multirow{4}{*}{Workload}
    & \textit{Parametric} & \textbf{Changing Predicate Distributions} \\
    \cline{3-3}
    &                  & \textbf{Varying Selectivity} \\
\cline{2-3}
    & \textit{Structural}       & \textbf{Modifying Query Structure}\\
    \cline{3-3}
    &                  & \textbf{Changing Payloads}\\
\hline
\end{tabular}
\caption{Hierarchical classification of drift operations.}
\label{tab:hierarchical-drift}
\end{table}

\section{Formalizing Drift}


In this section, we formalize data and workload drift by introducing a taxonomy of common drift types, along with their representative operations. Table~\ref{tab:hierarchical-drift} summarizes all categories covered in this work.

\subsection{Data Drift}

Data drift refers to changes in the cardinality or distribution of records within a database. We categorize it into four representative operations observed in benchmarks and research systems:

\begin{enumerate}[leftmargin=2em, topsep=0pt, itemsep=0pt, parsep=0pt]
    \item \textbf{Varying cardinality} refers to changes in the overall number of records in the dataset, typically scaled by a configurable factor (e.g., 0.5$\times$, 2$\times$, 10$\times$). This drift models data growth and is used to assess system scalability~\cite{sosd,TPCH,TPCDS}.

    \item \textbf{Updating cardinality} involves inserting or deleting records, commonly used to evaluate how secondary indices handle continuous updates~\cite{rtree,RSMI,r_star_tree}.
    While often described as update-intensive workloads~\cite{ULI_Evaluation}, we treat it as data drift due to the impact on the data.

    \item \textbf{Shifting column distributions} refers to changes in the statistical distribution of column values, such as increased skewness~\cite{PR_tree}. 
    This form of drift is common in spatial systems, where skewed distributions can evaluate robustness, and typically occurs without changes in dataset cardinality~\cite{RSMI,BMTree,LBMC}.

    \item \textbf{Injecting outliers} uses rare or extreme values to test system robustness under distributional anomalies~\cite{Poisoning_SVM}. 
    In PostgreSQL, such outliers can distort column statistics, leading the optimizer to misestimate selectivity and choose suboptimal plans~\cite{PostgresVacuum}.
    While this issue has gained attention in learned index research (e.g., data poisoning~\cite{Poisoning}), it remains largely overlooked in other database components.
    As machine learning components integrate more into database systems~\cite{NeurDB}, handling outliers will become increasingly critical.


\end{enumerate}

These operations form a practical foundation for modeling data drift. 
While we focus on these representative types, our goal is not to exhaustively enumerate all possible drift scenarios, but rather to capture the most common patterns.
These abstractions motivate the need for a formal definition, which we provide below.





\begin{definition}[Data Drift]
Let $\mathcal{D}_1$ and $\mathcal{D}_2$ denote two versions of a dataset over the same schema $\mathcal{S}$. 
We define data drift as a \textbf{significant change} in the statistical properties or volume of data between $\mathcal{D}_1$ and $\mathcal{D}_2$. 
It can be characterized along two primary subtypes:

\begin{itemize}[leftmargin=1em, topsep=0pt, itemsep=0pt, parsep=0pt]

    \item \textbf{Cardinality Drift:} A substantial change in the total number of records, i.e., $|\mathcal{D}_1|$ and $|\mathcal{D}_2|$ differ by more than a defined threshold $\alpha$ 
    (e.g., $\left| |\mathcal{D}_2| - |\mathcal{D}_1| \right| > \alpha \times |\mathcal{D}_1|$).

    \item \textbf{Distributional Drift:} A change in the column value distribution.
    Such changes can be categorized as (i) \emph{global}, where a divergence metric $\delta(\mathcal{D}_1, \mathcal{D}_2)$ exceeds a threshold $\epsilon$ (e.g., changes in skewness); or (ii) \emph{local}, where small-scale modifications (e.g., point injection) affect specific regions of the distribution but can substantially influence system behavior.
    
\end{itemize}
\end{definition}

\fix{
The thresholds $\alpha$ and $\epsilon$ in our definition are not parameters required by \ourtool. In practice, such thresholds are commonly defined by database systems or practitioners based on operational policies. 
For example, in Oracle, the deletion of 20\% of a table's rows is used as a threshold to trigger index rebuilding~\cite{oracle_rebuild}.
}

\begin{figure}[h]
\subfloat[Dataset $D_1$~\label{fig:data_drift_diff_stage_0}]   {
\includegraphics[width=0.15\textwidth]{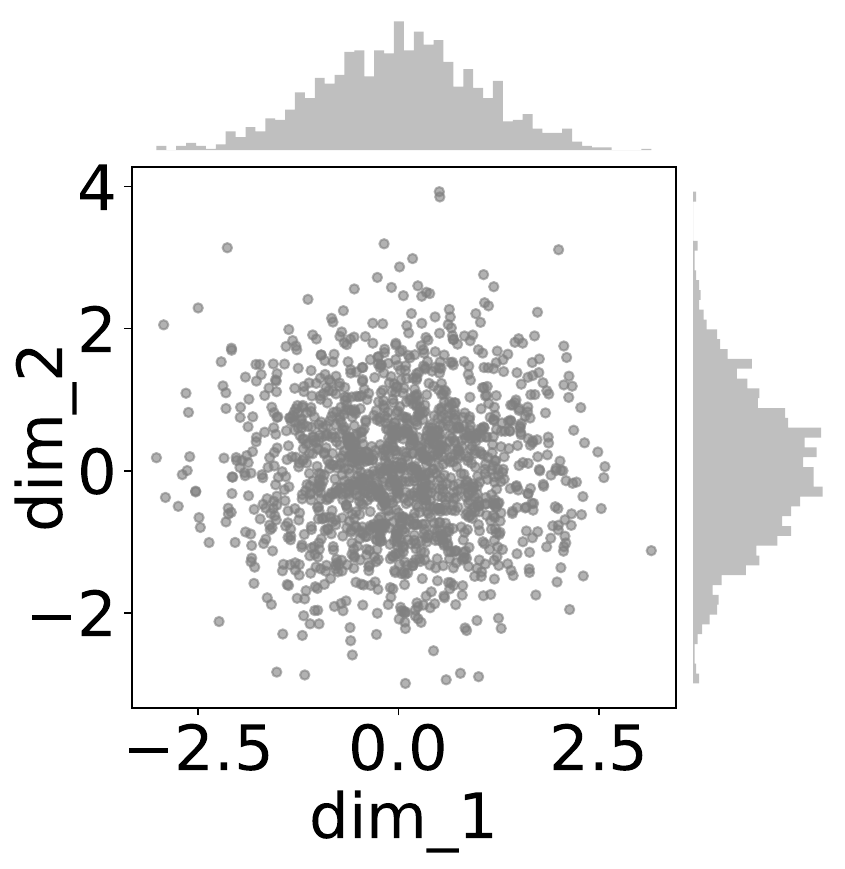}
}
\hspace{-0.5em}
\subfloat[Cardinality drift $D_2$~\label{fig:data_drift_diff_stage_1}]   {
\includegraphics[width=0.15\textwidth]{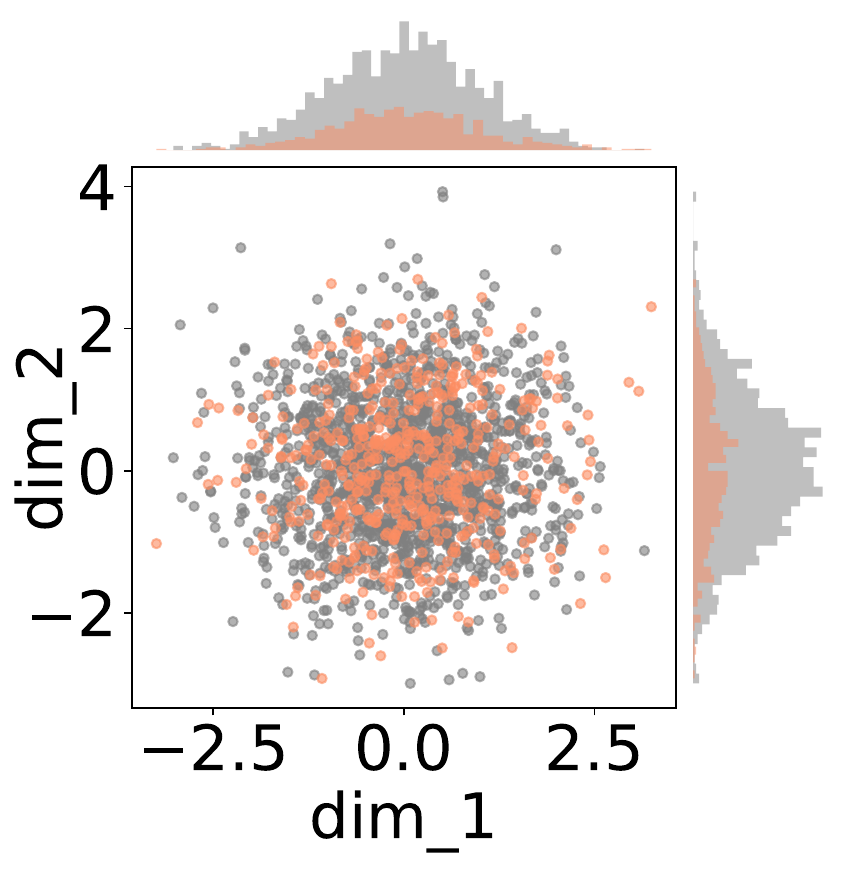}
}
\hspace{-0.5em}
\subfloat[Distributional drift $D_3$~\label{fig:data_drift_diff_stage_2}]   {
\includegraphics[width=0.15\textwidth]{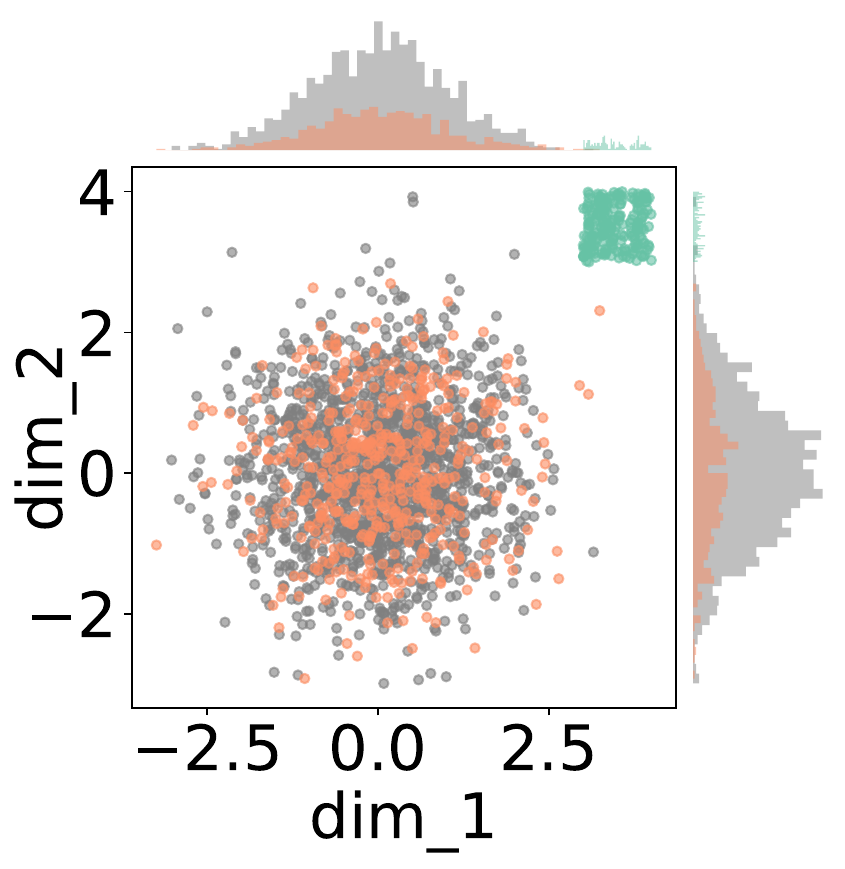}
}
\caption{Illustration of data drift.}
\label{fig:data_drift_diff_stage}
\end{figure}

\noindent \textbf{Example.} 
Figure~\ref{fig:data_drift_diff_stage} illustrates an example of progressive data drift.
In $D_1$, the data follows a stable normal distribution on two dimensions. In $D_2$, a new mode emerges (orange), where the data distribution partially overlaps with the original. 
In $D_3$, a second mode appears (green), shifted further in value space due to the injection of outliers. This reflects how real-world data can evolve through overlapping shifts.

\noindent \textbf{Discussion.} 
While we define each drift type individually, in practice, multiple types may co-occur between two time points. Our \ourtool\ supports such compound drift by allowing intermediate drifted datasets to be reused as inputs, enabling the simulation of multi-drift scenarios over time. However, since drift combinations are often case-specific and  domain-dependent~\cite{data_shift_ML}, we do not formalize drift operation composition in this paper.

\subsection{Workload Drift}
Workload drift refers to changes in the structure or statistical properties of queries executed against a database over time. 
We categorize workload drift into four common operations, each of which can significantly impact query processing behavior:

\begin{enumerate}[leftmargin=2em, topsep=0pt, itemsep=0pt, parsep=0pt]
    \item \textbf{Changing predicate distributions} reflects shifts in the statistical distribution of predicates over time. This directly affects query optimizers~\cite{Bao,Balsa} and learned indices~\cite{LMSFC,LBMC,BMTree} that rely on historical access patterns.
    
    \item \textbf{Varying selectivity} arises when queries from the same logical template exhibit varying predicate ranges~\cite{DBA_bandit,BMTree,RSMI}. Such drift commonly occurs in response to changing analytical demands (e.g., broader time ranges) and leads to different join strategies or plan choices.
    
    \item \textbf{Modifying query structure} captures structural changes in query templates, such as modified predicates or join conditions~\cite{DBA_bandit,Bao,Neo,Balsa,are_learned_ready}. These shifts can trigger re-optimization or impact index usage.

    \item \textbf{Changing payloads} refers to changes in the set of projected columns. While 
    it modifies the query structure, we treat it as a separate drift type due to its distinct impact on I/O cost and column scan behavior~\cite{redbench,are_learned_ready,QUILTS}.
\end{enumerate}

\fix{
While payload, predicate, and join are orthogonal, we group predicate and join due to their impact on cardinality, which influences plan generation. 
In contrast, payload defines the output schema and has relatively limited influence.}
We now present a formal definition of workload drift based on these observations.




\begin{definition}[Workload Drift]
A workload $W$ is defined as a distribution $P(W_\tau(\theta))$ over queries instantiated from a parameterized template $\tau$, where $\theta \in \Theta$ denotes the parameter-generating operator.
Let $\text{Sel}(W_\tau(\theta))$ denote the total selectivity of $W$.
Workload drift is characterized by the following subtypes:

\begin{itemize}[leftmargin=1em, topsep=0pt, itemsep=0pt, parsep=0pt]
    \item \textbf{Parametric Drift:} Two workloads $W_{1,\tau}$ and $W_{2,\tau}$ are instantiated from the same template $\tau$, but differ in their parameter distribution. 
    That is, for some $\theta_1, \theta_2 \in \Theta$,
    $\delta(P(W_{1,\tau}(\theta_1)),P(W_{2,\tau}(\theta_2))) > \epsilon$ or $|\text{Sel}(W_{1,\tau}(\theta_1)) - \text{Sel}(W_{2,\tau}(\theta_2))| > \alpha \times \text{Sel}(W_{1,\tau}(\theta))$.
    

    \item \textbf{Structural Drift:} Two workloads $W_{\tau_1}$ and $W_{\tau_2}$ are instantiated from
    templates $\tau_1$ and $\tau_2$, respectively, with structural differences in predicates, joins, or payloads.
\end{itemize}

\end{definition}

\fix{We omit the detailed explanation of $\delta$,
$\alpha$, and $\theta$, as they have been discussed in the context of data drift.}

\begin{figure}[h]
\subfloat[Workload $W_1$~\label{fig:query_shape_stage_0}]   {
\includegraphics[width=0.15\textwidth]{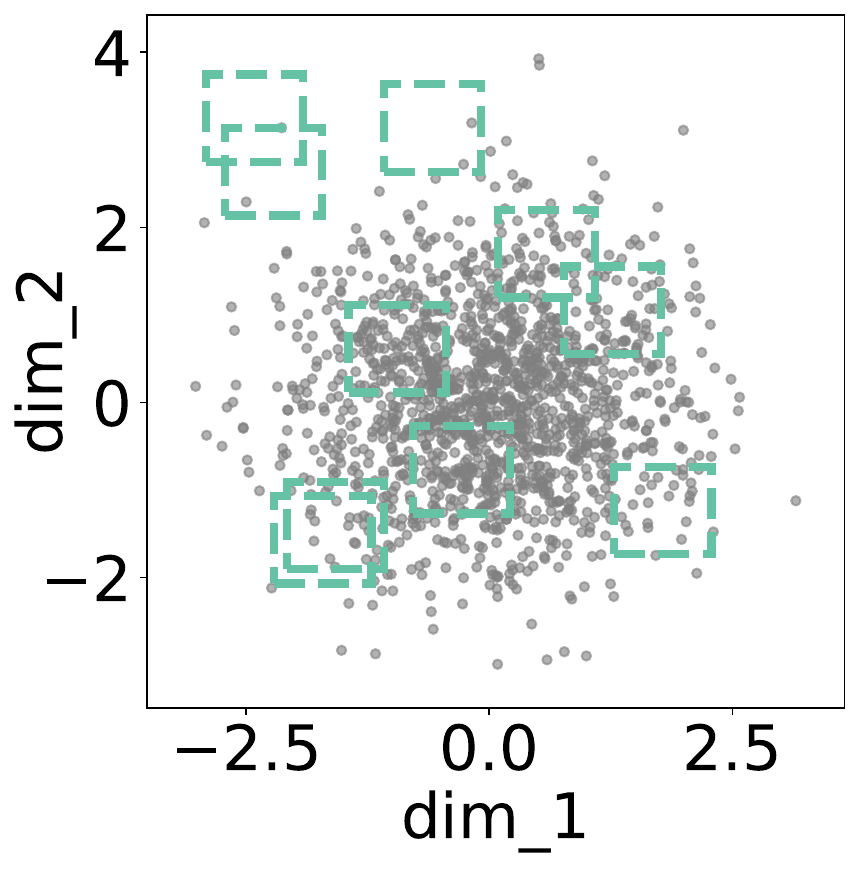}
}
\hspace{-0.5em}
\subfloat[Parametric drift $W_2$~\label{fig:query_shape_stage_1}]   {
\includegraphics[width=0.15\textwidth]{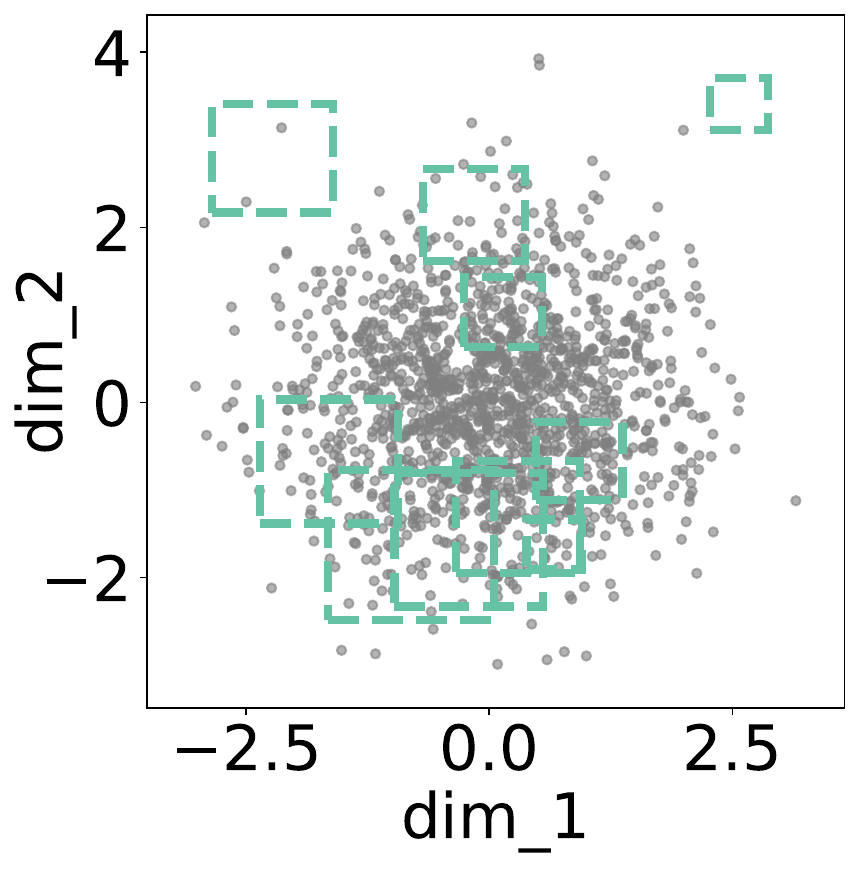}
}
\hspace{-0.5em}
\subfloat[Structural drift $W_3$~\label{fig:query_shape_stage_2}]   {
\includegraphics[width=0.15\textwidth]{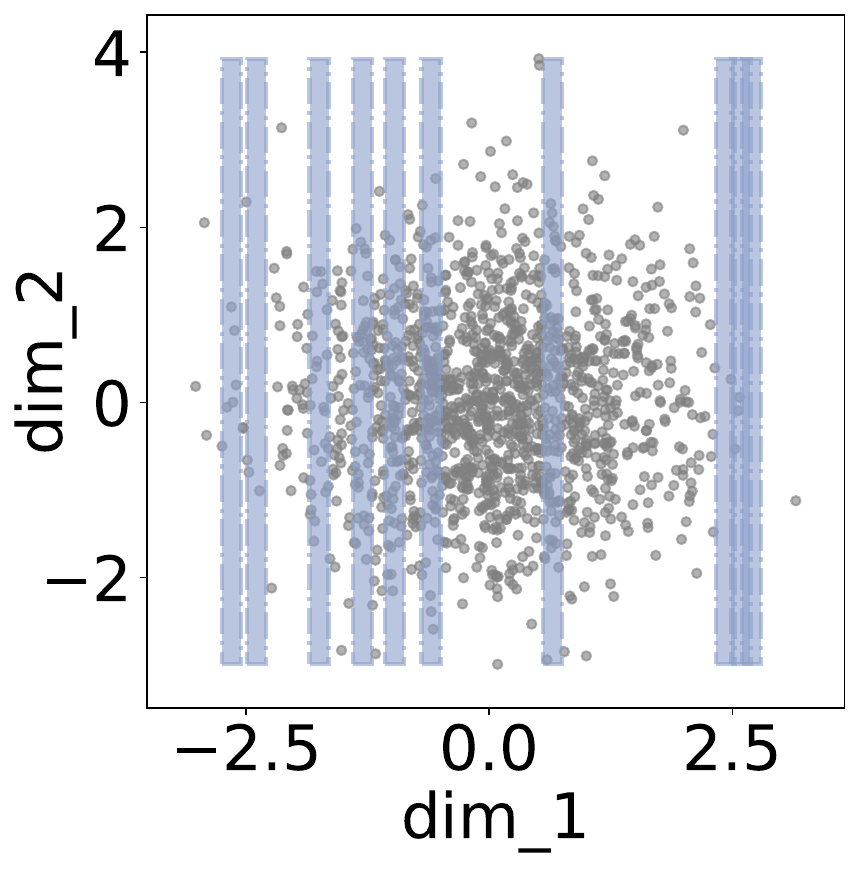}
}
\caption{Illustration of workload drift across time.}
\label{fig:query_shape_stage}
\end{figure}

\noindent \textbf{Example.}  
Figure~\ref{fig:query_shape_stage} shows workload drift. Each green dashed box represents a query’s predicate region over two dimensions.
In $W_1$, queries are uniformly distributed with identical predicate bounds, modeling a stable workload.
In $W_2$, the predicate ranges shift and vary slightly (i.e., \textit{parametric drift}).
In $W_3$, queries (blue) remove their predicate on \texttt{dim\_2}, resulting in a broader scan (i.e., \textit{structural drift}).
Such evolution mirrors real-world analytics, where user interests and data exploration patterns shift gradually.

\subsection{Temporal Drift Patterns}
We model temporal drift as the evolution of data or workloads over time, typically following non-stationary patterns such as bursts, trends, or repeats. 
Rather than giving a new definition, we adopt the taxonomy in Sibyl~\cite{Sibyl}. 
In particular, we use four representative temporal patterns: uniform, periodic, trend, and long-tail.

\begin{figure}[h]
  \centering
  \includegraphics[width=0.45\textwidth]{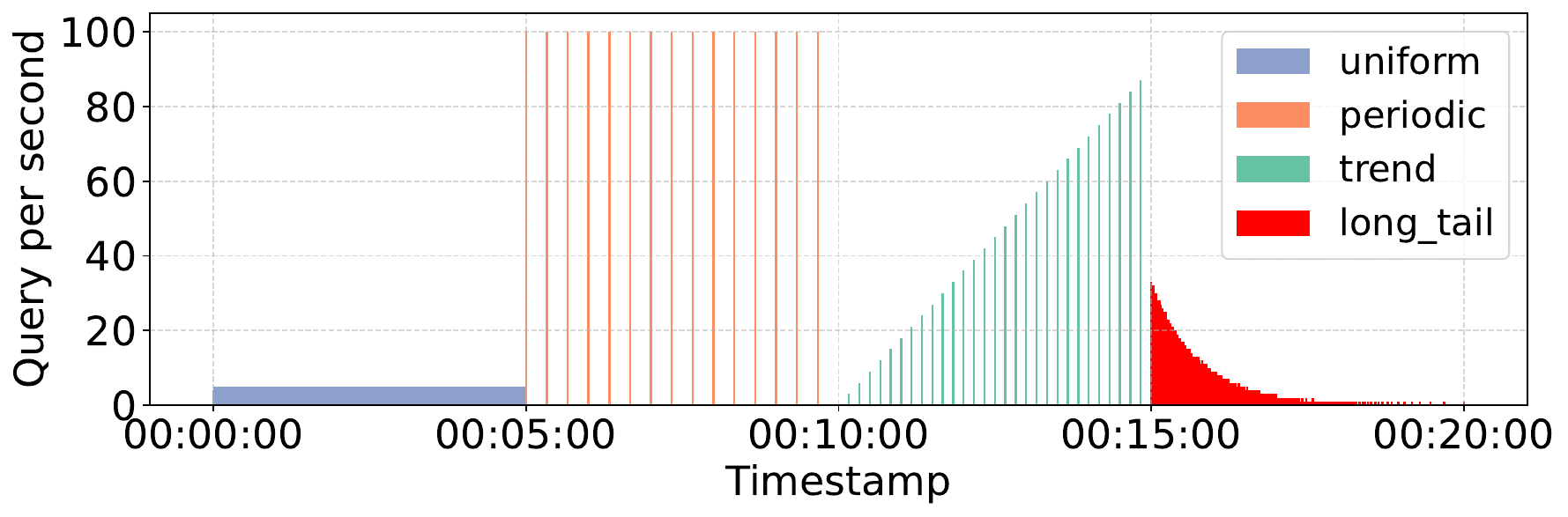}
  \caption{Four temporal drift patterns.}
  \label{fig:case_study_frequency}
\end{figure}

\noindent \textbf{Example.}  
Figure~\ref{fig:case_study_frequency} illustrates four timestamp patterns over a five-minute window, with 1,500 query instances.
In the uniform pattern, timestamps are evenly spaced at a constant rate of five queries per second.
The periodic pattern generates 100 queries every 20 seconds, forming periodic bursts.
The trend pattern increases the arrival rate linearly over time.
In the long-tail pattern, most queries arrive early, with frequency dropping sharply afterward.

\noindent \textbf{Discussion.}
These patterns can be applied independently or in combination with data and workload drift, enabling more expressive and customizable drift scenarios. 
For example, query timestamps can be generated per instance to simulate realistic query streams with temporal variation. 
\textit{Repetitive query instances} are common in real-world cases~\cite{redbench} and can be assigned with periodic timestamps. 
Additionally, data updates, such as deletions and insertions, can be aligned with specific time intervals to emulate read-heavy or write-heavy patterns~\cite{ULI_Evaluation}. 
These capabilities allow users to construct complex temporal-evolving scenarios.

\section{System Design and Extensibility}


\subsection{System Design}
\begin{figure}[h]
  \centering
  \includegraphics[width=0.45\textwidth]{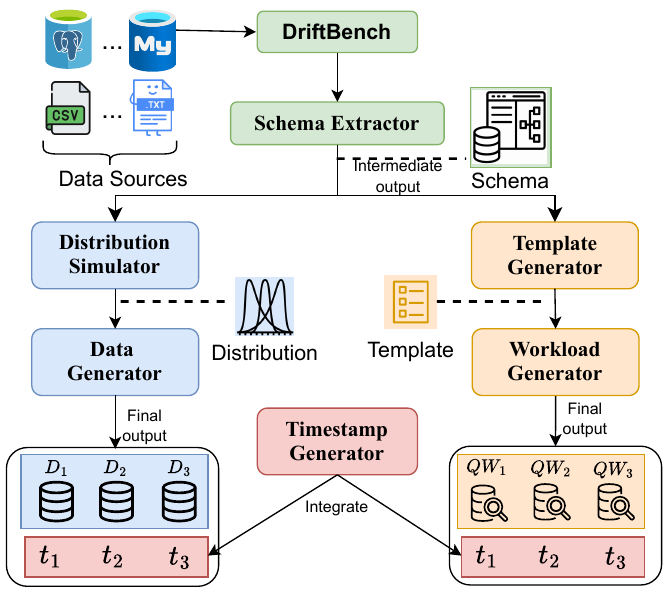}
  \caption{Architecture of \ourtool.}
  \label{fig:framework}
\end{figure}

As shown in Figure~\ref{fig:framework}, \ourtool\ takes various data sources as input and extracts detailed schema information. Intermediate results, including schema, distributions, and templates, are used to generate drifted data and workloads.

\ourtool\ has six key components:

\begin{itemize}[leftmargin=1em, topsep=0pt, itemsep=0pt, parsep=0pt]
    \item \textbf{Schema extractor} parses the input data source and extracts a schema containing column-level metadata, such as logical types (e.g., numeric, categorical, datetime), sample values, value ranges, and empirical percentiles, histogram bounds.
    It supports both file-based and relational sources, with current implementations for CSV and PostgreSQL, two of the most commonly used formats in benchmarking and prototyping.
    
    \item \textbf{Distribution simulator} applies controlled transformations to simulate data drift, including value skew, outlier injection, cardinality variation, and selective deletion. 
    Drift can be column-specific and parameterized by intensity.
    
    \item \textbf{Data generator} synthesizes new rows based on column-level statistics inferred from the schema, using type-aware methods such as KDE~\cite{KDE} for numerics and frequency-based sampling for categoricals. 
    The module supports cardinality scaling, row deletions, and insertions.
    \fix{For multi-table cases, we generate relation-consistent data by jointly sampling join keys and ensuring that referencing tables include values compatible with the base tables.}

    \item \textbf{Template generator} generates parameterized query templates over the extracted schema. 
    Each template defines a query structure, including filter predicates, payloads, and optional joins. 
    The generator supports both single-table and multi-table workloads by leveraging column-level statistics and join candidates.
    \fix{
    The key difference is that multi-table templates include \emph{join} clauses. 
    While joins often follow foreign key relationships in schemas, such constraints may be absent in benchmarks like TPC-DS~\cite{TPCDS}. 
    To address this, 
    we implement a join inference mechanism based on column name similarity (e.g., using \texttt{SequenceMatcher}). 
    
}


    \item \textbf{Workload generator} instantiates query templates by sampling concrete predicate values from specified distributions (e.g., uniform, normal, Zipfian). 
    This module transforms abstract logical templates to executable queries.

    \item \textbf{Timestamp generator} can generate timestamps following defined patterns. 
    It simulates non-stationary query arrival processes by controlling inter-arrival intervals, enabling realistic modeling of time-evolving workloads.
\vspace{-1em}
\end{itemize}

\subsection{Extensibility}
\ourtool\ is designed to be modular and extensible, making it easy to support new data sources, drift operations, 
and temporal patterns for future use cases. We briefly describe four main extensions:

\begin{enumerate}[leftmargin=2em, topsep=0pt, itemsep=0pt, parsep=0pt]
    \item \textbf{Pluggable Input Sources.} 
    While our current implementation supports CSV and PostgreSQL, additional backends (e.g., MySQL, dat files, and Pandas DataFrame) can be supported by implementing the same interface.

    \item \textbf{Custom Drift Operations.}
    Drift types are modeled as composable transformations over data or query templates. Users can define new operations by specifying transformation logic, target columns, and parameter controls. 



    \item \textbf{Template Generator Hooks.}
    Template generation is driven by user-defined constraints (e.g., number of predicates, target selectivity). The generator can easily adapt to existing query templates, e.g., TPC-H~\cite{TPCH} and TPC-DS~\cite{TPCDS}.

    \item \textbf{Temporal Pattern Plugins.}
    Time-evolving behavior is modeled independently of data or workload types. 
    New temporal modes (e.g., exponential decay) can be plugged into the timestamp generator without altering query logic.
\end{enumerate}

 \vspace{-0.5em}
\section{Case Studies}

We use \ourtool\ to generate data and workload drift (Sections~\ref{subsec:data_drift} and~\ref{subsec:workload_drift}) on the \texttt{census} dataset~\cite{census_dataset,are_learned_ready}, and evaluate their impact on cardinality estimation (Section~\ref{subsec:cardinality_estimation}).




 \vspace{-0.5em}
\subsection{Data Drift}~\label{subsec:data_drift}
We select two representative attributes from the \texttt{census} dataset: one numeric (\texttt{age}) and one categorical (\texttt{workclass}). For categorical features, we show only the top 5 most frequent categories. 

\begin{figure}[h]
\includegraphics[width=0.225\textwidth]{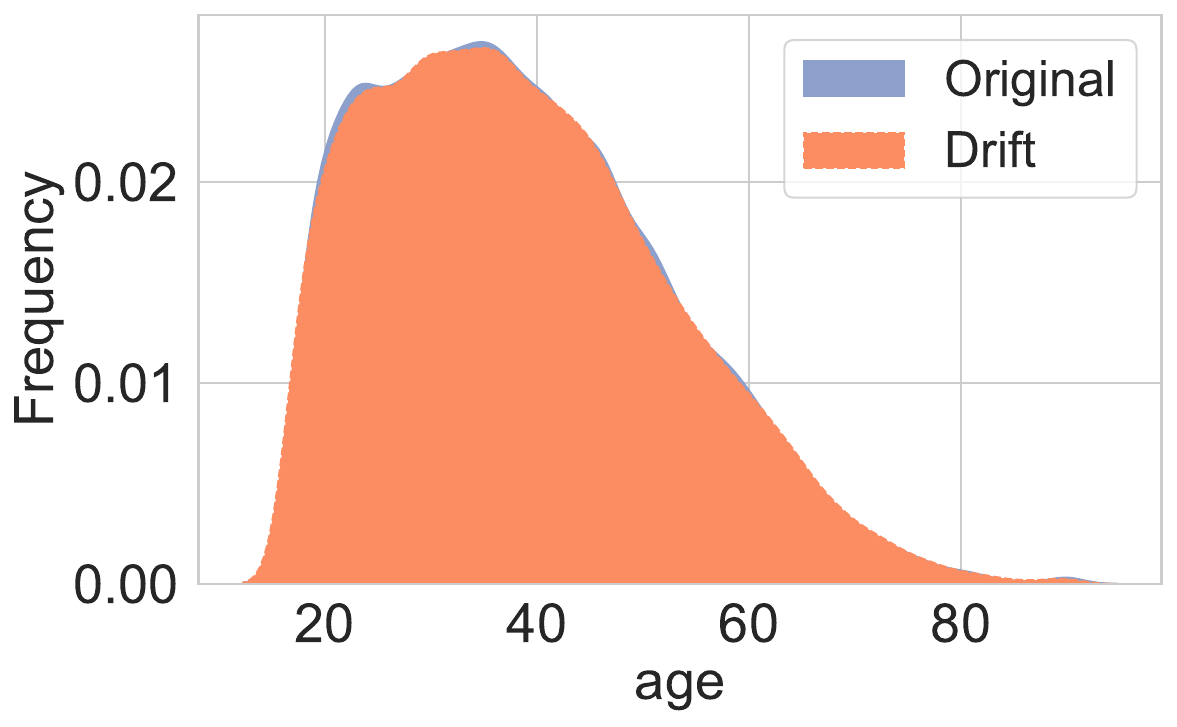}
\includegraphics[width=0.225\textwidth]{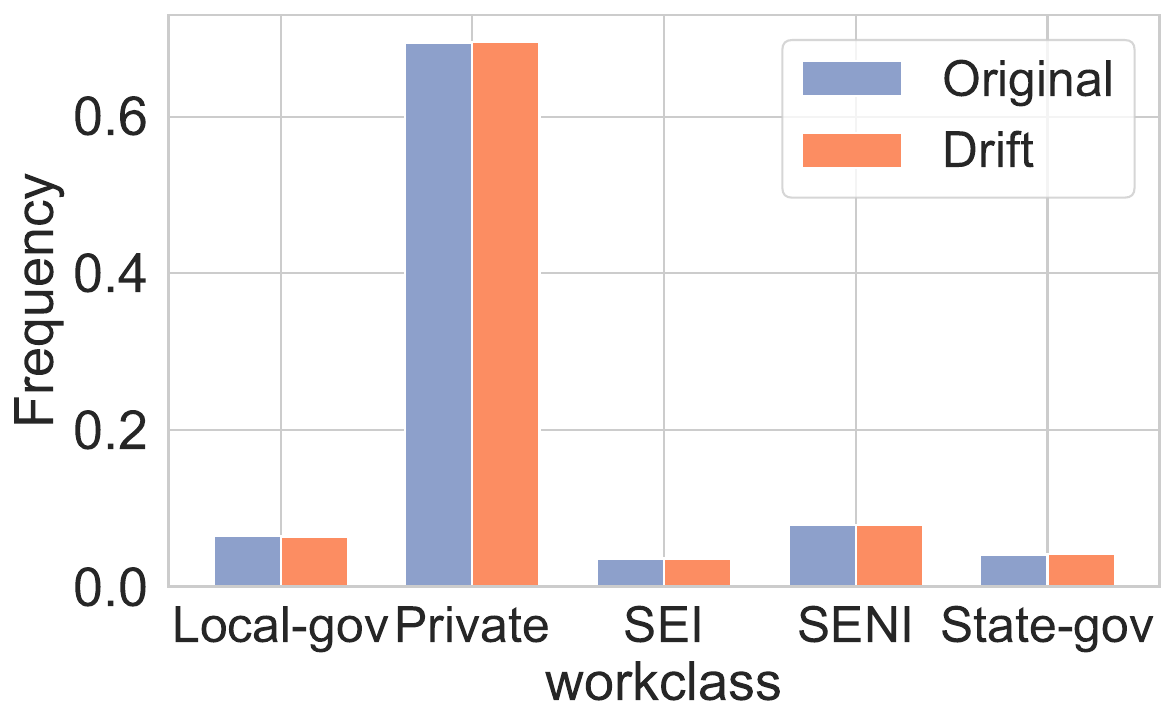}
  \caption{Data distributions under cardinality variation.}
\label{fig:case_study_cardinality_scale}
\end{figure}

\subsubsection{Varying Cardinality} 
We simulate cardinality variation by generating a dataset of the same size as the original, while preserving attribute distributions.
Figure~\ref{fig:case_study_cardinality_scale} shows the distribution consistency between the original and scaled datasets, 
confirming that attribute-level patterns are preserved during cardinality drift.




\subsubsection{Updating Cardinality}\label{subsubsec:updating_cardinality}
Cardinality updates reflect row-level changes from insertions and deletions. While insertions reuse scaling, deletions act on existing data. To simulate realistic deletions, 10\% of the \texttt{census} dataset is sampled proportionally to its original distribution, avoiding artificial bias.




\begin{figure}[h]
\includegraphics[width=0.225\textwidth]{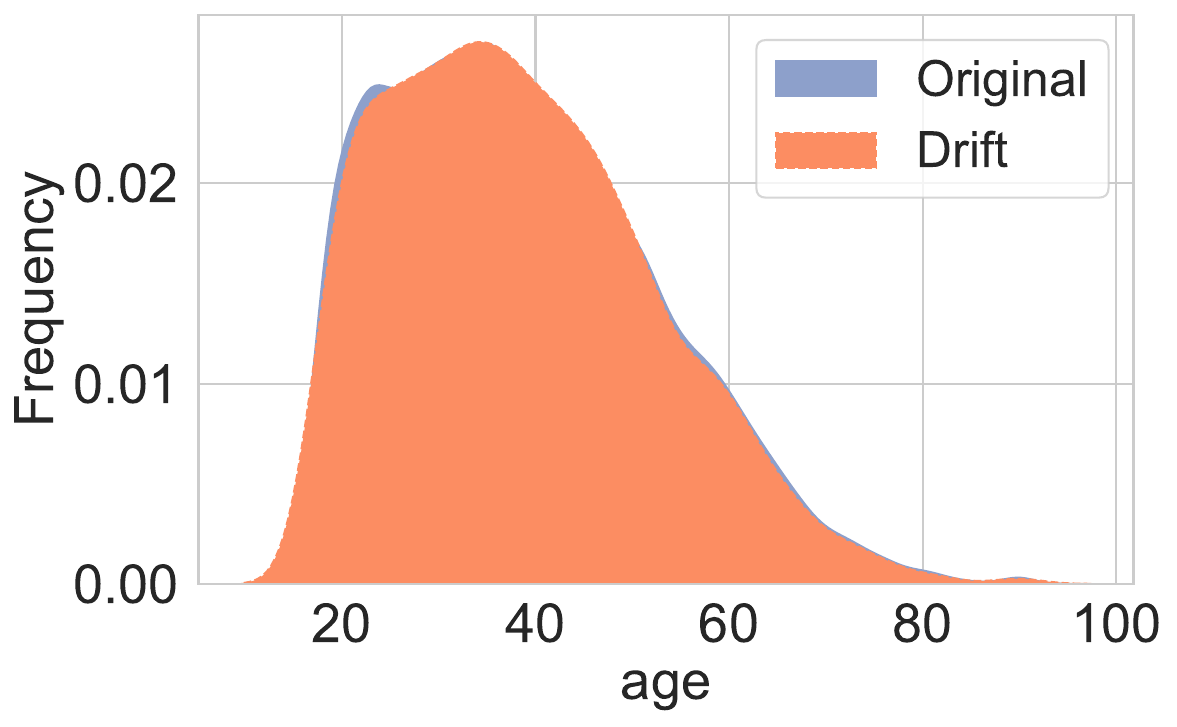}
\includegraphics[width=0.225\textwidth]{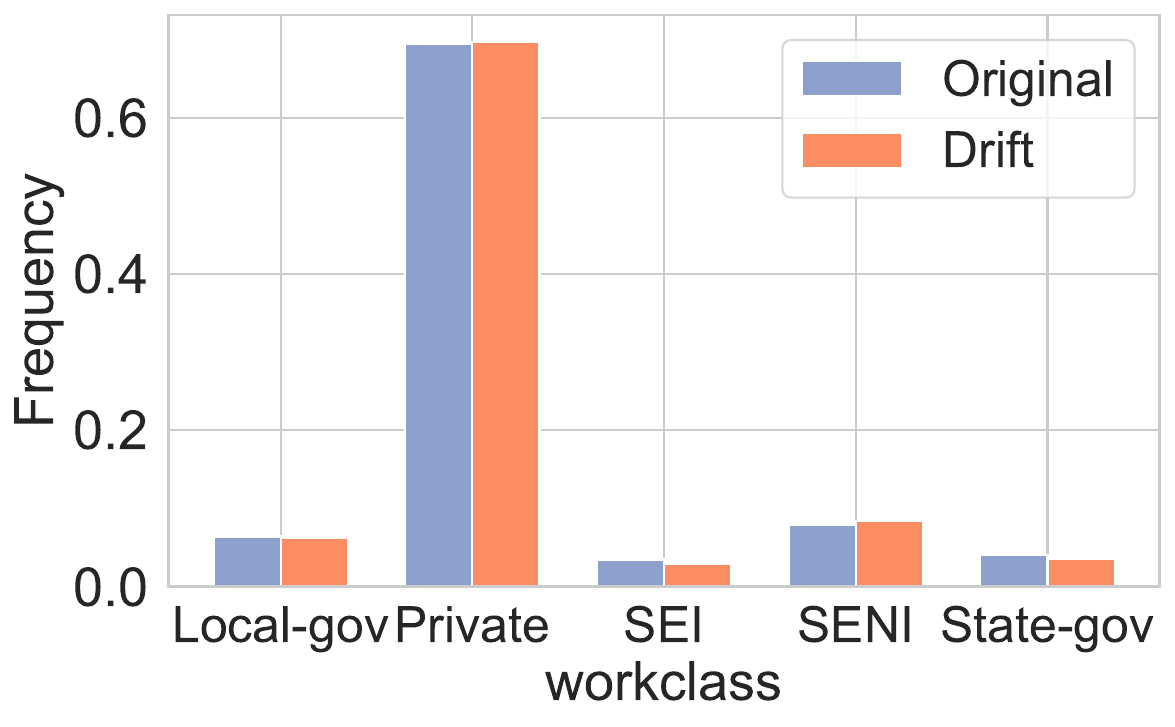}
  \caption{Data distributions under cardinality updates.}
\label{fig:case_study_cardinality_delete}
\end{figure}


 \vspace{-0.5em}
\subsubsection{Shifting Column Distributions}\label{subsubsec:shifting_Distribution}
\begin{figure}[h]
\includegraphics[width=0.225\textwidth]{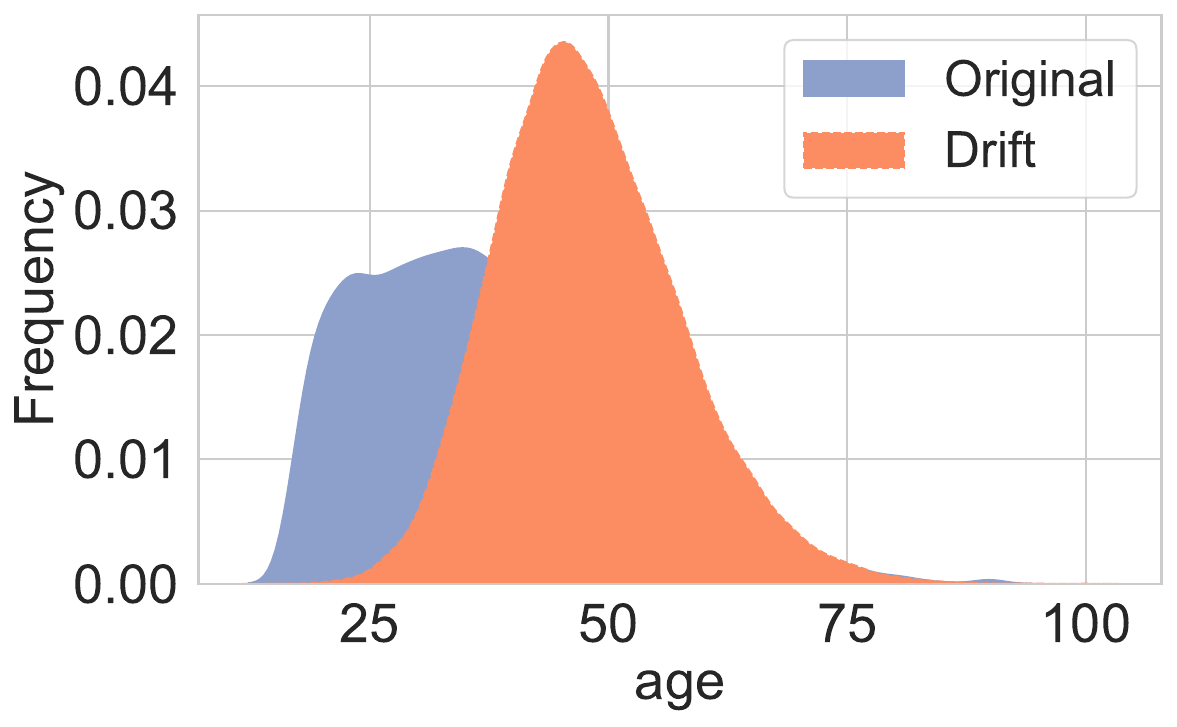}
\includegraphics[width=0.225\textwidth]{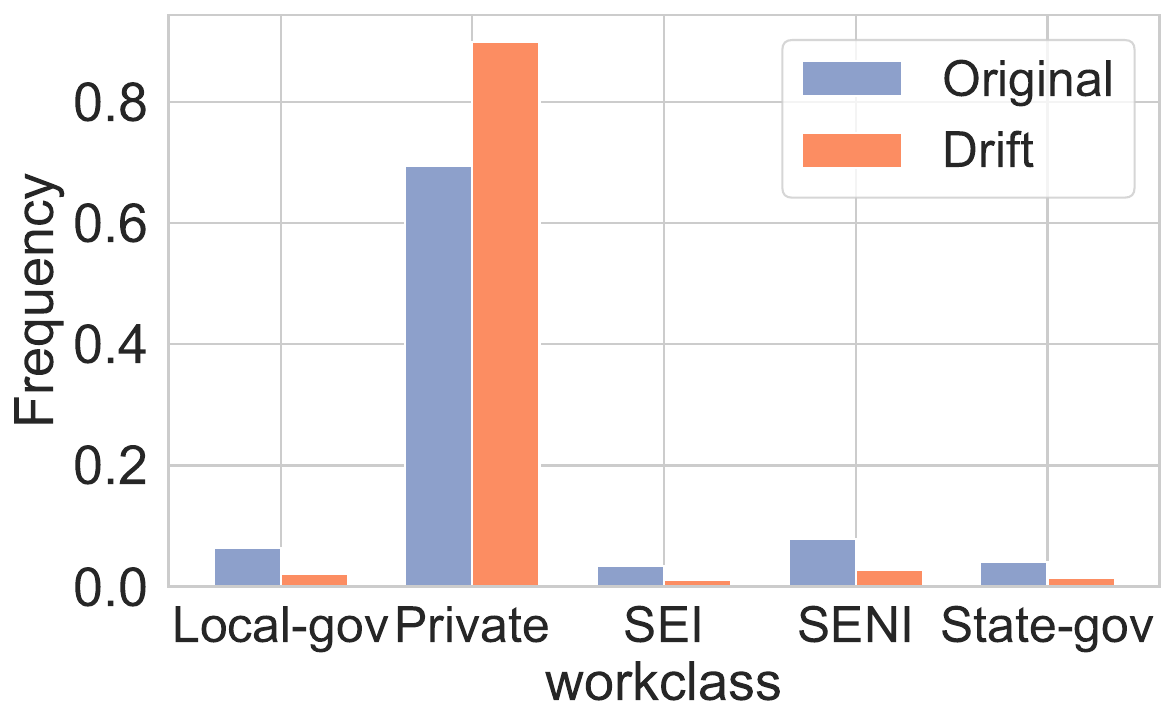}
  \caption{Data distributions under column skew.}
\label{fig:case_study_census_original_skew}
\end{figure}

We simulate drift by replacing original columns with more skewed distributions. 
For numeric attributes, we generate values that preserve the mean and standard deviation but increase skewness toward one side. 
For categorical attributes, we upweight the most frequent categories to increase skew, amplifying the dominance of popular values. Figure~\ref{fig:case_study_census_original_skew} illustrates these changes for representative columns.




 \vspace{-0.5em}
\subsubsection{Injecting Outliers} 
\begin{figure}[h]
  \centering
  \includegraphics[width=0.48\textwidth]{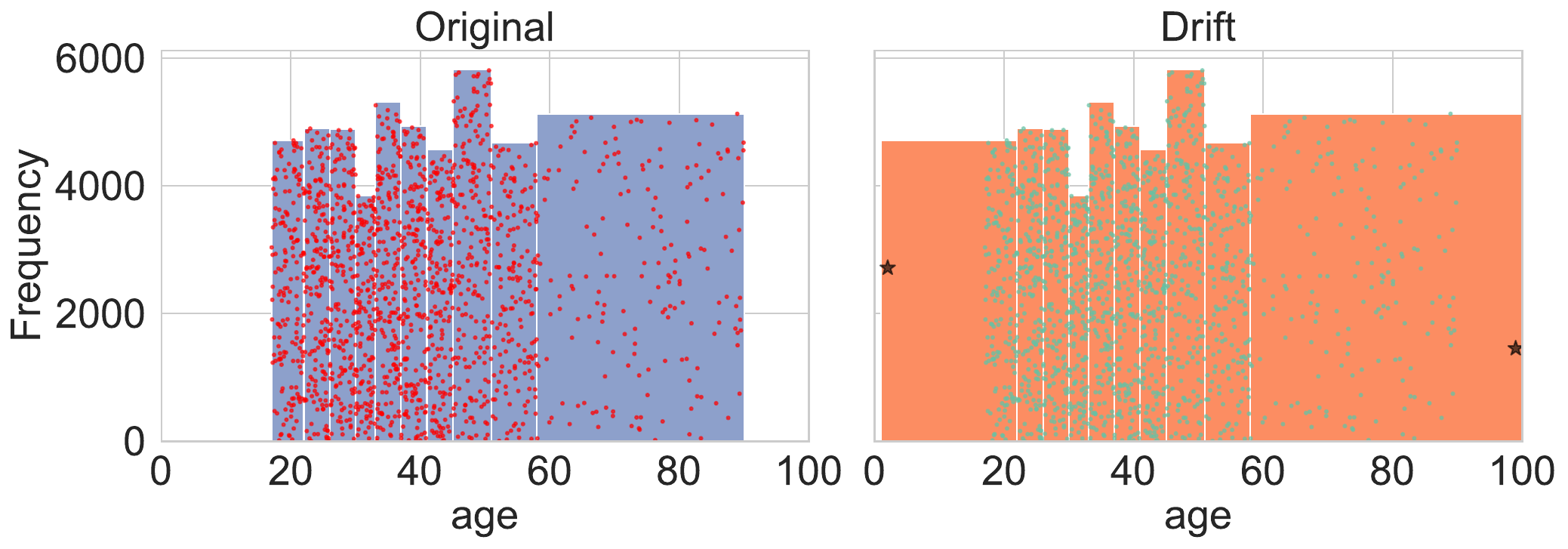}
  \caption{Data histograms under outliers injection.}
  \label{fig:case_study_census_original_outlier}
\end{figure}

In systems like PostgreSQL, inserting outliers can confuse the query planner and lead to poor performance. As shown in Figure~\ref{fig:case_study_census_original_outlier}, we inject two outliers \texttt{age=1} and \texttt{age=100} into \texttt{census}, where most ages range from 18 to 90. 
These values may not be well-represented in the statistics, leading to suboptimal plans.
PostgreSQL relies on histogram bounds for selectivity estimation, which are only updated after \texttt{ANALYZE} or \texttt{VACUUM}.
Therefore, these values can be missing from the optimizer's statistics, causing poor selectivity estimates and suboptimal plans.


\vspace{-0.5em}
\subsection{Workload Drift}~\label{subsec:workload_drift}
We use the \texttt{age} column from the \texttt{census} dataset to illustrate different forms of workload drift in conjunction with timestamps.

\subsubsection{Changing Predicate Distributions}\label{subsubsec:changing_predicate} 
\begin{figure}[h]
  \centering
  \includegraphics[width=0.45\textwidth]{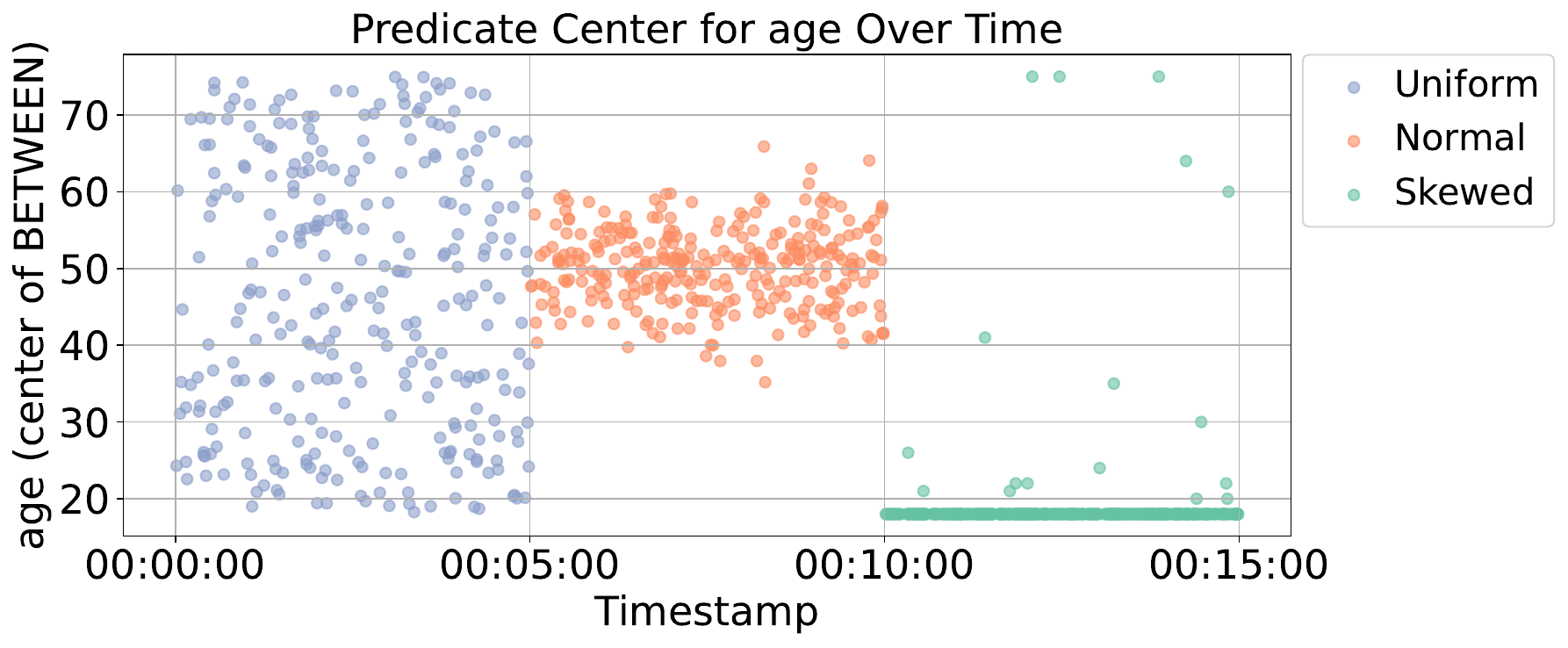}
  \caption{Predicate center drift}
  \label{fig:case_study_predicate_center}
\end{figure}

We demonstrate predicate shift with three workloads by varying the value distribution of the \texttt{age} column across three timestamp groups, each uniformly distributed within five minutes.
As shown in Figure~\ref{fig:case_study_predicate_center}, the query predicates follow uniform, normal, and skewed distributions, respectively, reflecting increasing levels of locality and skew.

\vspace{-0.5em}
\subsubsection{Varying Selectivity}\label{subsubsec:varying_selectivity}
\begin{figure}[h]
  \centering
  \includegraphics[width=0.45\textwidth]{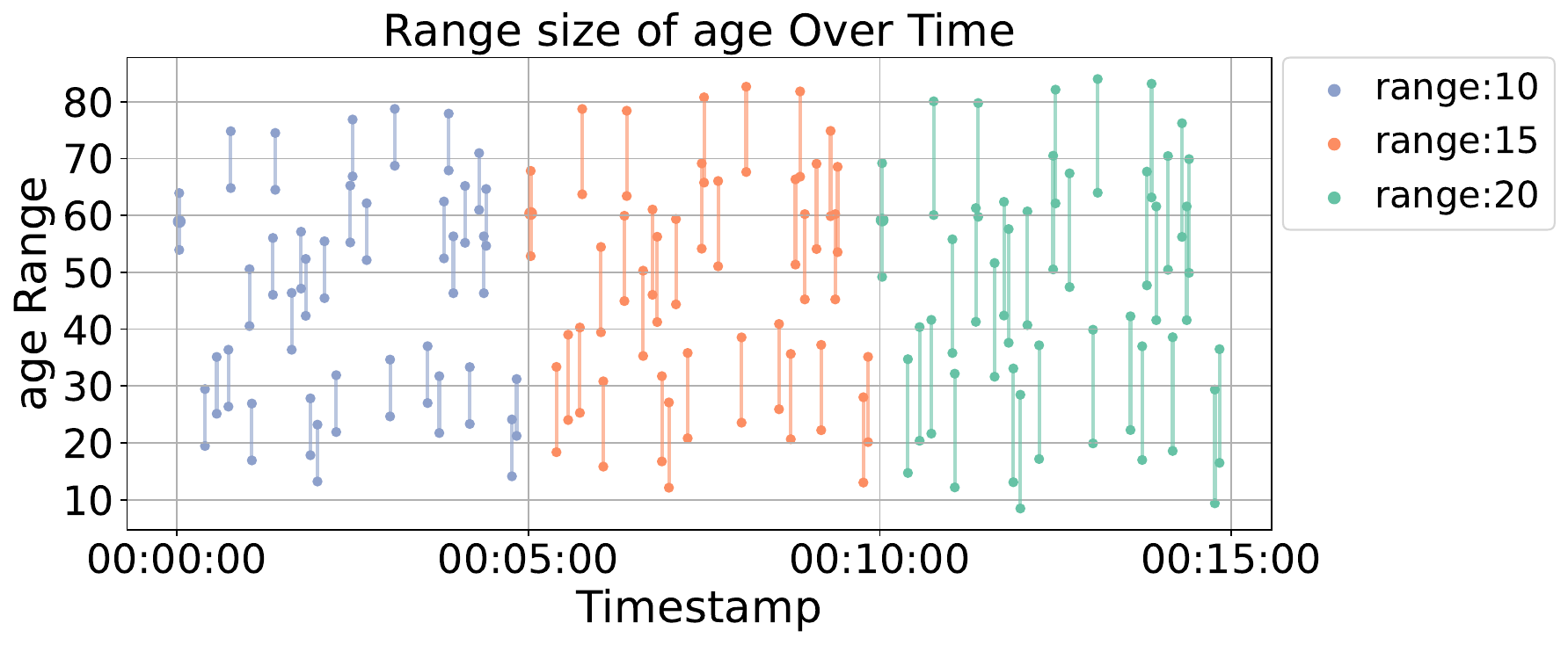}
  \caption{Selectivity drift}
  \label{fig:case_study_predicate_range_size}
\end{figure}

While we do not directly estimate query selectivity, we simulate selectivity drift by progressively expanding the predicate range over time. 
As shown in Figure~\ref{fig:case_study_predicate_range_size}, there are three workloads, and we increase the range of the \texttt{age} predicate linearly (from 10 to 20) across three groups of timestamps.

\vspace{-0.5em}
\subsubsection{Structural Drift} 
\begin{figure}[h]
  \centering
  \includegraphics[width=0.4\textwidth]{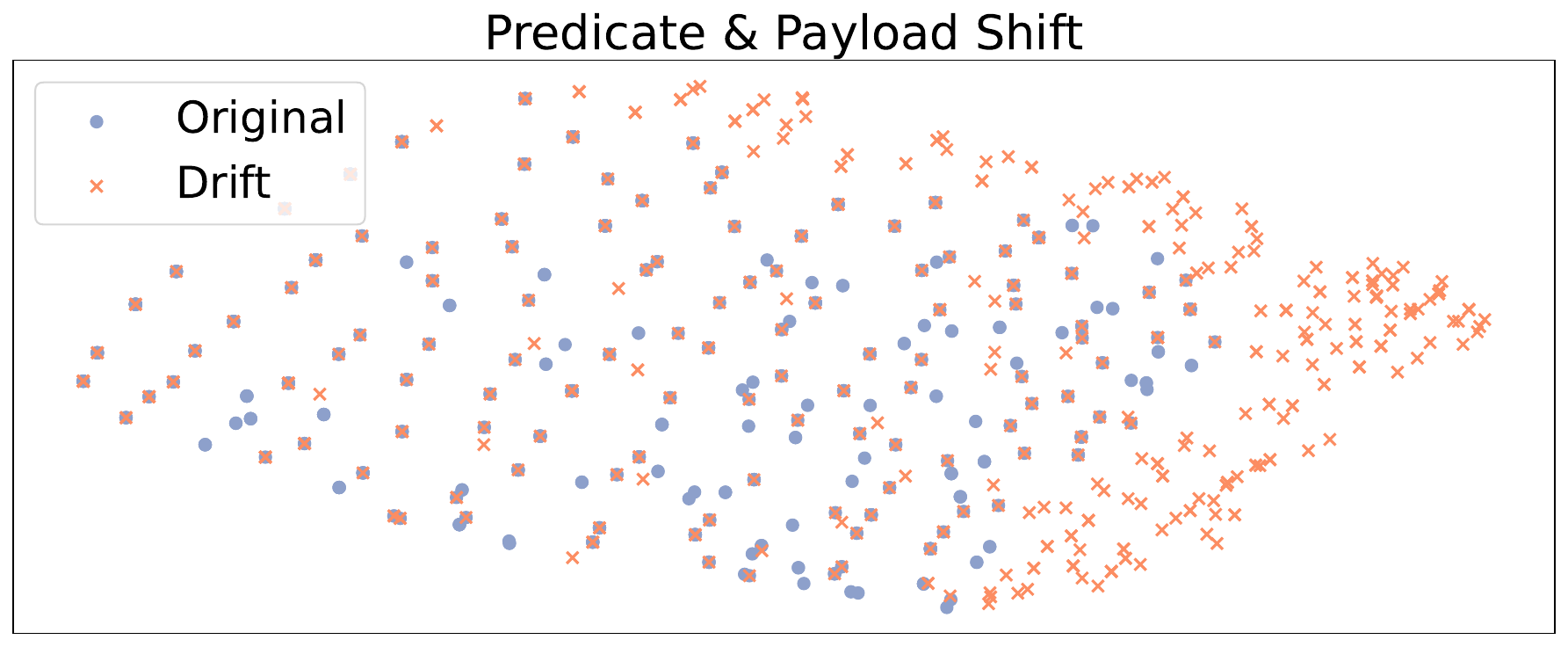}
  \caption{t-SNE for structural drift.}
  \label{fig:tsne_pred_payload}
\end{figure}

We group these two operations (\textit{modifying query structure} and \textit{changing payloads}) together, as logical mutations often co-occur with changes in the number of predicates, joins, and other clauses. 
These changes are structural and difficult to visualize directly.
To capture their overall effect, we extract high-level features from each query template and project them using t-SNE (t-distributed stochastic neighbor embedding).
Figure~\ref{fig:tsne_pred_payload} visualizes the difference between two sets of query templates: one generated with at most 5 predicates and 6 projected columns, and another with larger limits of 7 and 8, respectively. The resulting clusters reflect how query structures evolve under logical and payload drift.

\vspace{-0.5em}
\subsection{Evaluating Estimator Behavior under Drift}\label{subsec:cardinality_estimation}
\begin{figure}[h]
\includegraphics[width=0.45\textwidth]{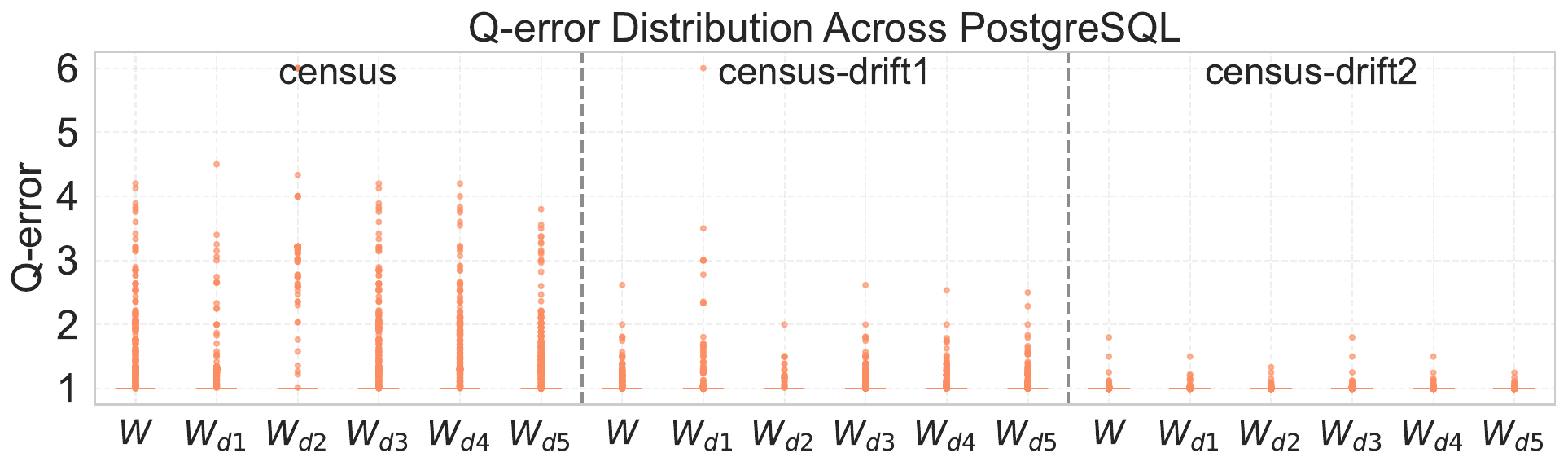}
\includegraphics[width=0.45\textwidth]{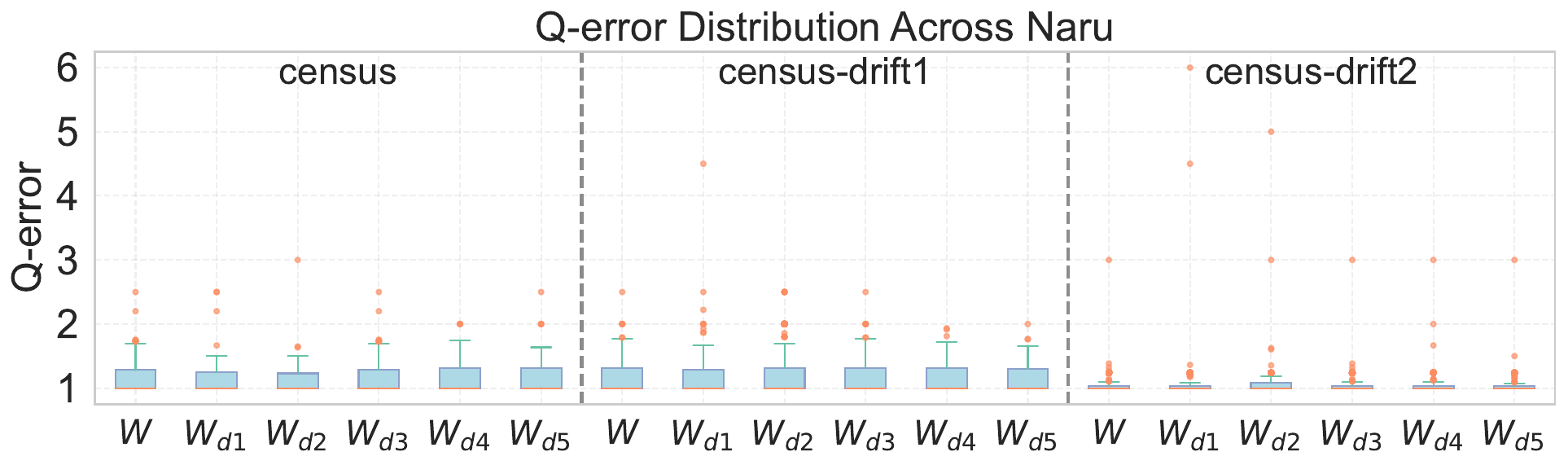}
\includegraphics[width=0.45\textwidth]{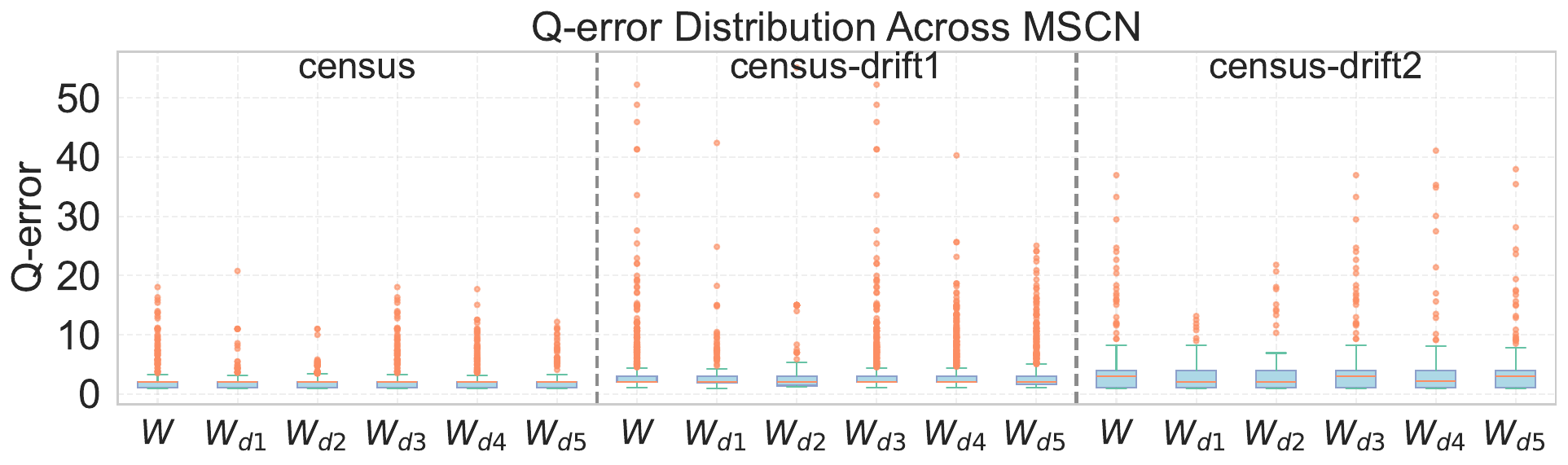}
  \caption{ Q-error patterns under controlled drift reveal estimator-specific behavioral differences.}
\label{fig:case_study_ce}
\end{figure}

\fix{
We use this experiment to examine how different types of drift affect the behavior of cardinality estimators. 
Rather than comparing absolute accuracy, we focus on how estimators respond to systematic changes in data and workloads.

We evaluate three representative estimators: PostgreSQL (rule-based), Naru~\cite{Naru} (data-driven), and MSCN~\cite{MSCN} (data- and query-driven). Using the \texttt{census} dataset, we generate two data variants: \texttt{census-drift1} (cardinality updates) and \texttt{census-drift2} (distributional skew), described in Sections~\ref{subsubsec:updating_cardinality} and~\ref{subsubsec:shifting_Distribution}. 
For each dataset, we apply six workloads: $W$ is the uniform workload from Section~\ref{subsubsec:changing_predicate}. $W_{d1}$ and $W_{d2}$ correspond to normal and skewed distributions, respectively, and $W_{d3}$ to $W_{d5}$ vary in selectivity, as described in Section~\ref{subsubsec:varying_selectivity}.

As shown in Figure~\ref{fig:case_study_ce}, each estimator exhibits distinct behavior under data and workload drift. 
For PostgreSQL, the average Q-error remains relatively stable across datasets. However, the number and spread of outliers vary noticeably across data versions (\texttt{census}, \texttt{census-drift1}, and \texttt{census-drift2}). 
Moreover, within each dataset, different workloads lead to distinct outlier patterns.
Naru, being purely data-driven, shows consistent Q-error across workloads within each dataset, but its performance changes significantly in \texttt{census-drift2}, where the learned model no longer aligns with the drifted data.
MSCN, which incorporates both data and query features, reflects both forms of drift: its Q-error varies noticeably across both datasets and workloads, showing sensitivity to both data changes and predicate variations.

These findings underscore the need to evaluate database components under controlled drift scenarios. 
\ourtool\ enables such analysis across a wide range of components. 
Although this study focuses on cardinality estimation, we plan to extend the methodology to indexing, caching, and query planning.
}

\vspace{-1em}
\section{Related Work}
Classic benchmarks like TPC-H~\cite{TPCH} and TPC-DS~\cite{TPCDS} target analytical workloads over static schemas and datasets, while TPC-C~\cite{TPCC} and OLTPBench~\cite{oltpbench} focus on transactional throughput. 
DSB~\cite{DSB} adds dynamic workloads to TPC-DS, but supports only its original schema.
More specialized efforts, such as the JOB~\cite{job}, are designed for cardinality estimation and use real-world statistics and query templates, but still operate on static workloads.

RedBench~\cite{redbench} takes a step forward by incorporating real-world query patterns sampled from Amazon Redshift~\cite{RedShift}, capturing query and distribution drift. However, it remains limited to replaying existing queries and does not support controlled data drift generation or workload generation across schemas.
In contrast, \ourtool\ adopts a generative approach, synthesizing both datasets and workloads with precise control over data and query drift. 
If RedBench is a search engine for existing queries, \ourtool\ is like a generative model for data and workload drift.
To the best of our knowledge, \ourtool\ is the first tool to support both data and workload drift.

\vspace{-0.8em}
\section{Conclusion}
In this paper, we address core gaps in the definition and generation of data and workload drift within the database community.
We categorize each type of drift into four representative scenarios and provide formal definitions to standardize their usage.
To support drift simulation, we introduce \ourtool, a lightweight framework for generating synthetic data and workloads under controlled drift.
\ourtool\ also supports integration with temporal drift patterns, enabling more realistic and expressive testing.
We demonstrate the effectiveness of our approach through case studies.
As future work, we plan to extend \ourtool\ to support non-tabular systems such as vector, key-value, and graph databases.

\vspace{-0.8em}
\section*{Acknowledgments}
\vspace{-0.2em}
We gratefully acknowledge support from the Australian Research
Council Discovery Early Career Researcher Award DE230100366,
and Google Foundational Science 2025 fund.

\vspace{-0.8em}
\bibliographystyle{ACM-Reference-Format}
\bibliography{references}

\end{document}